\documentclass{aa} 
\usepackage{multirow}
\usepackage[normalem]{ulem}

\usepackage{booktabs}
\usepackage{multirow}
\usepackage{array}
\usepackage{graphicx}
\usepackage{txfonts}
\usepackage{footmisc}
\usepackage{capt-of}

\usepackage[colorlinks=true,linkcolor=blue,citecolor=blue,urlcolor=blue]{hyperref}

\usepackage[switch]{lineno}

\begin{document} 

%\linenumbers
\title{A matter of class and flare}
\subtitle{Flare productivity in solar active regions and 
power-law constraints on the flare frequency distribution}

 \author{A.G.M.\ Pietrow\inst{1} 
     }

 \institute{\inst{1}Leibniz-Institut für Astrophysik Potsdam (AIP), An der Sternwarte 16, 14482 Potsdam, Germany\\
       \email{apietrow@aip.de}\\
      }

\date{Draft: compiled on \today}

\abstract
{Solar active regions are the source of virtually all significant space weather events, therefore the relationship between their observable properties and flare production is important to characterize.}
{The flare productivity of solar active regions is analyzed across solar cycles 23--25 as a function of magnetic complexity, sunspot area, and solar-cycle phase. Additionally, the power-law index of the peak flux frequency distribution is estimated.}
{The GOES soft X-ray event reports matched with daily Solar Region Summaries (SRS) are used to create a robust active region catalog. This catalog is supplemented with the 50-year HEK flare catalog and the newly released STIX flare catalog from Solar Orbiter to estimate the power-law index.}
{Magnetically complex active regions ($\beta\gamma$, $\beta\delta$, $\beta\gamma\delta$) produce the majority of M- and X-class flares while comprising only $\sim$13\% of all daily active-region observations, showing that larger flares disproportionately occur in larger area sunspots. All three catalogs can be represented with a consistent power-law index of $\alpha = 2.02 \pm 0.01$ for the peak flux frequency distribution. However, a class-by-class analysis of the C/X and M/X flare ratios reveals that the distribution steepens significantly towards higher energies. Fitting this steepening as a lognormal tail and extrapolating two decades beyond the X-class threshold predicts a recurrence time of $370^{+292}_{-159}$~yr for an X100-class (`Carrington-type') superflare, compared to only $\sim$26~yr from a naive single power-law extrapolation.}
{While $\alpha \approx 2$ provides a useful global description of the flare frequency distribution, the steepening area-productivity relation and the divergence between C/X and M/X derived slopes both point to a deficit of the most energetic flares relative to a single power-law extrapolation. This implies that extreme events such as Carrington-type flares are considerably rarer than a fixed power law would predict, with our extrapolated recurrence time in much closer agreement with independent literature estimates.}

\keywords{Sun: flares -- Sun: activity -- Sun: X-rays -- Sun: corona -- methods: statistical}

 \maketitle
%
%________________________________________________________________

\section{Introduction}

The most severe disruptions to the heliosphere are caused by solar flares and the subsequent coronal mass ejections (CMEs) \citep[e.g.][]{Wedemeyer2025}. The accurate modeling of these events is a crucial yet largely unresolved aspect of space weather prediction with many directions of research \citep[e.g.][]{Gallagher2002, Bloomfield2012, McCloskey2016,Kontogiannis2023,Kontogiannis2024b,Kontogiannis2024}. One such direction explores the flare productivity of solar active regions (ARs) \citep[for a review, see][]{Toriumi2019}.

The relationship between AR properties and flare production has been studied for nearly a century, beginning with \citet{Giovanelli1939}, who first examined flare occurrence in relation to sunspot size and type. Since then, our understanding has been greatly refined by continuous synoptic monitoring from the Geostationary Operational Environmental Satellite \citep[GOES,][]{Woods2024}, the Solar and Heliospheric Observatory \citep[SOHO,][]{Domingo1995}, and later the Solar Dynamics Observatory \citep[SDO,][]{Pesnell2012}. 

It is now well established that flare productivity correlates with the magnetic complexity of the host AR, with complex ARs being disproportionately responsible for the most energetic flares \citep{Kunzel1960, Sammis2000, Gao2014, Kontogiannis2024, Krivova2026}. The dependence on sunspot area has also been documented \citep{Lee2012, Eren2017}, though it is generally found to be secondary to the role of magnetic complexity \citep{Sammis2000}. 

Beyond the properties of individual regions, the aggregate energy distribution of solar flares has been a topic of sustained interest since the pioneering work of \citet{Drake1971}. This distribution is commonly described by a power law, which can be extrapolated to predict the occurrence of much rarer superflares \citep{Vasilyev2024}, as well as the contribution of micro- and nanoflares to coronal heating \citep{Hudson1991}.

Despite this extensive body of work, most studies have either focused on individual AR parameters in isolation, considered only a single solar cycle, or treated the flare population as a whole without distinguishing between flare energy classes. This work aims to address this gap by combining the 
30-year synoptic flare and AR catalogs from the Space 
Weather Prediction Center (SWPC). Additionally, the frequency distribution of flares themselves will be investigated. 

\section{Observations}
\begin{figure*}
\centering
\includegraphics[width=\textwidth]{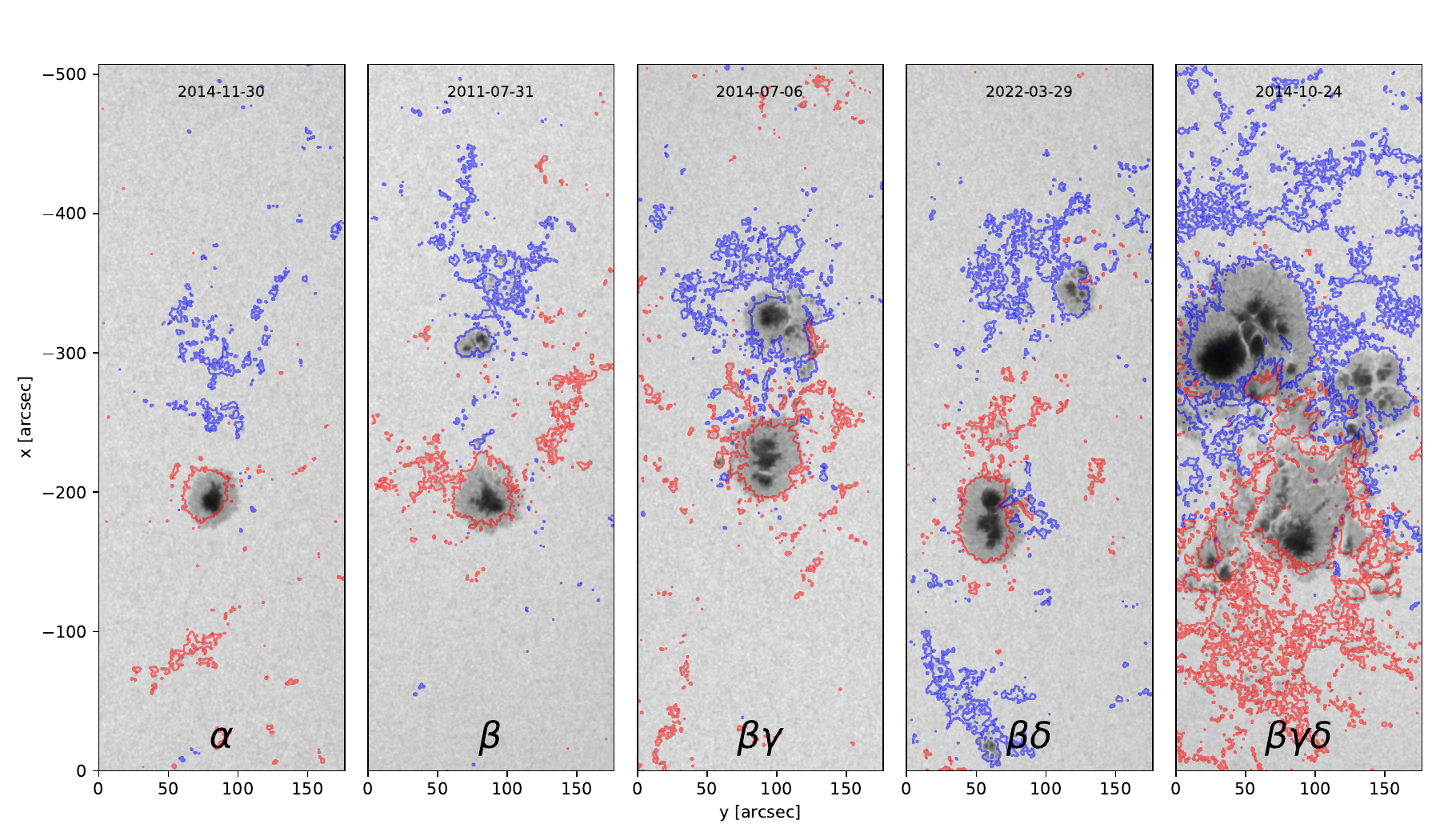}
\caption{Example active regions for each Hale magnetic classification observed by SDO/HMI.  The background shows the continuum intensity and the overlaid contours indicate leading (red) and trailing (blue) magnetic polarity from the line-of-sight magnetogram.}
\label{fig:classes}
\end{figure*}
We use three complementary datasets for our flare statistics research, which are described below. Only one of these sets has a matching AR catalog, with the other two being used primarily to compare the obtained results on flare frequencies.

\subsection{GOES data}
The Geostationary Operational Environmental Satellite \citep[GOES;][]{hanser1996} missions have monitored solar flares in soft X-rays (1--8\,\AA) since 1975, and have since been succeeded by multiple generations, with GOES-18 and 19 currently in operation. Each new satellite overlaps in time with its predecessor, enabling simultaneous observations of the same flare events and thereby allowing inter-satellite cross-calibration and the derivation of consistent calibration factors \citep[e.g.][]{Hudson2024}. Each flare's peak X-ray flux is expressed on the logarithmic GOES classification scale \citep{Baker1970}, with strengths classified by letter (A, B, C, M, or X) representing an order-of-magnitude increase in flux and a numerical multiplier within each class. In this work, we assume that these calibrations are correct and accurate, and do not further process the data.

Flare information, including event lifetimes (start, peak, and end times), class, and location, is obtained from the Heliophysics Events Knowledgebase \citep[HEK;][]{Hurlburt2012}. This dataset spans 50 years and contains close to 100,000 individual flare entries, providing the long baseline needed for the flare frequency analysis. However, the HEK data do not contain information on the corresponding ARs that gave rise to these flares. 

A second dataset, providing the AR linkage needed for our productivity analysis, is derived from the SWPC daily event reports, which combine GOES X-ray measurements with ground-based H$\alpha$ observations from the Solar Observing Optical Network \citep{lucas2025}. SWPC forecasters assign each flare to the nearest active region, tying it to a National Oceanic and Atmospheric Administration (NOAA) region number that is in turn characterised in terms of magnetic classification, sunspot area, and temporal evolution through the daily Solar Region Summaries\footnote{\url{https://www.ngdc.noaa.gov/stp/space-weather/swpc-products/daily_reports/}} (SRS). This dataset spans 30 years. 

In general, active regions are designated a NOAA number if they meet one or more of the following criteria \citep{SWPC2012, pietrowthesis}:
\begin{enumerate}
    \item The region has a sunspot group with spot class C or larger according to the \citealt{McIntosh1990} classification.
    \item Two or more reports confirm the presence of a class A or B spot group, preferably 12 hours apart.
    \item The region produces a solar flare.
    \item The region is bright in H$\alpha$ and exceeds 5 heliographic degrees in either latitude or longitude.
\end{enumerate}

The daily SRS bulletin lists provide a list of ARs found on the disk. Each entry contains a NOAA region number, heliographic position, corrected sunspot area in millionths of the solar hemisphere (MSH), Zurich/McIntosh morphological classification \citep[][]{McIntosh1990}, and a modified Mount Wilson (Hale) magnetic classification \citep[][]{AFWA2013_SOON} which is based on the traditional classification \citep[][]{Hale1919} with the addition of $\delta$-spots \citep[][]{Kunzel1965} and a merger of several subclasses.

In this work, the modified Hale classification as given by \citet{Jaeggli2016} is used, as this system is more widely adopted in the space weather community\footnote{See e.g.\ \url{https://www.spaceweatherlive.com/en/help/the-magnetic-classification-of-sunspots.html}} than the McIntosh classification \citep{McIntosh1990}. The individual classes are defined below and shown in Fig.~\ref{fig:classes}.
\begin{itemize}
    \item[$\alpha$] A unipolar sunspot group.
    \item[$\beta$] A bipolar sunspot group with positive and negative polarities, separated by a clear and simple division.
    \item[$\delta$] A spot containing umbrae of opposite polarity enclosed within a single penumbra. This type is only defined to be used in the classifications below, and is not individually assigned to ARs by the NOAA.
    \item[$\beta$$\gamma$] A bipolar sunspot group that is sufficiently complex that no single line can be drawn between spots of opposite polarity.
    \item[$\beta$$\delta$] A sunspot group with an overall $\beta$ magnetic configuration that contains one or more $\delta$ sunspots.
    \item[$\beta$$\gamma$$\delta$] A sunspot group with a $\beta$$\gamma$ magnetic configuration that contains one or more $\delta$ sunspots.
\end{itemize}
ARs can be more coarsely subdivided into simple ($\alpha$ and $\beta$) and complex ($\beta\gamma$, $\beta\delta$, $\beta\gamma\delta$) regions. A total of 45\,220 entries are contained within this dataset, of which 86.8\% fall into the simple category and the remaining 13.2\% are classified as complex. Following \citet{Jaeggli2016}, each entry is treated as an independent event expressed in units of `AR-days', reflecting the fact that the statistics are recorded on a daily basis.

Flares without reported coordinates, typically those occurring near or beyond the limb, are assigned a default position of $(x,y)=(0,0)$ in the event files, and are included in general statistics but excluded from region-related analysis.
From the event reports, 52\,189 flare events are extracted, 33\,494 (64\%) of which are matched with ARs (see table~\ref{tab:hale_stats}).

All datasets in this work are taken `as is' with the assumption that all corrections between successive GOES generations have been properly applied.

\begin{table*}[h]
\caption{Distribution of AR-days, split into flaring and non-flaring populations, along with associated flare strengths. Flaring is defined as at least one XRA/XFL-classified flare (any GOES class) on that AR-day. Percentages in the flaring/non-flaring columns are relative to the total AR-days in that row. Percentages in the AR-days total column are relative to the full sample. Percentages in the flare columns are relative to the total number of flares in that class.}
\label{tab:hale_stats}
\centering
\resizebox{\textwidth}{!}{%
\begin{tabular}{|l|l|r|r|r!{\vrule width 1pt}r|r|r|r|r|r|}
\hline
\multirow{2}{*}{Type} & \multirow{2}{*}{Class} & \multicolumn{3}{c!{\vrule width 1pt}}{AR-days} & \multicolumn{6}{c|}{Flares} \\
\cline{3-11}
 & & \multicolumn{1}{c|}{Flaring} & \multicolumn{1}{c|}{Non-flaring} & \multicolumn{1}{c!{\vrule width 1pt}}{Total} & \multicolumn{1}{c|}{A} & \multicolumn{1}{c|}{B} & \multicolumn{1}{c|}{C} & \multicolumn{1}{c|}{M} & \multicolumn{1}{c|}{X} & \multicolumn{1}{c|}{Total} \\
\hline
\multirow{2}{*}{Simple}
  & $\alpha$ & 1529 (\phantom{0}10.8\%) & 12638 (\phantom{0}89.2\%) & 14167 (\phantom{0}31.3\%) & 4 (\phantom{00}0.1\%) & 1198 (42.6\%) & 1488 (53.0\%) & 111 (\phantom{0}4.0\%) & 9 (\phantom{00}0.3\%) & 2810 (\phantom{0}8.4\%) \\
  & $\beta$  & 6419 (\phantom{0}25.6\%) & 18674 (\phantom{0}74.4\%) & 25093 (\phantom{0}55.5\%) & 9 (\phantom{00}0.1\%) & 5045 (33.1\%) & 9210 (60.4\%) & 941 (\phantom{0}6.2\%) & 38 (\phantom{00}0.2\%) & 15243 (45.5\%) \\
\hline
\multirow{3}{*}{Complex}
  & $\beta\gamma$ & 2651 (\phantom{0}64.1\%) & 1482 (\phantom{0}35.9\%) & 4133 (\phantom{00}9.1\%) & 1 (\phantom{00}0.0\%) & 1591 (19.2\%) & 5879 (70.8\%) & 796 (\phantom{0}9.6\%) & 37 (\phantom{00}0.4\%) & 8304 (24.8\%) \\
  & $\beta\delta$ & 266 (\phantom{0}71.1\%) & 108 (\phantom{0}28.9\%) & 374 (\phantom{00}0.8\%) & 0 (\phantom{00}0.0\%) & 143 (14.9\%) & 646 (67.2\%) & 158 (16.4\%) & 14 (\phantom{00}1.5\%) & 961 (\phantom{00}2.9\%) \\
  & $\beta\gamma\delta$ & 1227 (\phantom{0}84.4\%) & 226 (\phantom{0}15.6\%) & 1453 (\phantom{00}3.2\%) & 2 (\phantom{00}0.0\%) & 576 (\phantom{0}9.3\%) & 4260 (69.0\%) & 1208 (19.6\%) & 130 (\phantom{00}2.1\%) & 6176 (18.4\%) \\
\hline
\multicolumn{2}{|l|}{Total} & 12092 (\phantom{0}26.7\%) & 33128 (\phantom{0}73.3\%) & 45220 (100.0\%) & 16 (\phantom{00}0.0\%) & 8553 (25.5\%) & 21483 (64.1\%) & 3214 (\phantom{0}9.6\%) & 228 (\phantom{00}0.7\%) & 33494 (100.0\%) \\
\hline
\end{tabular}%
}
\end{table*}

\subsection{STIX data}
The GOES catalog is supplemented with the flare list from the Spectrometer/Telescope for Imaging X-rays (STIX; \citealt{Krucker2020}) aboard Solar Orbiter (SO; \citealt{Muller2020}). This list\footnotemark[3] is an updated version of the data released in \citet{Xiao2023}, and contains 25\,218 flares observed between February 2021 and February 2025. Of these, 13\,855 are classified as above C-class according to their STIX-derived GOES-equivalent flux and are used in the power-law analysis below (Sect.~\ref{sec:powerlaw}); the remainder are B-class or fainter. These flares are not matched to ARs. Unlike the GOES-derived datasets, STIX is observed by a different instrument in a different orbit, making it a truly independent check on the flare frequency distribution derived from the GOES-based catalogs.

\section{Results and discussion}
In this section, the AR and flare data are analyzed under a range of different segmentations in order to characterize the flaring behavior of the Sun more comprehensively.

\begin{figure*}
\centering
\includegraphics[width=\textwidth]{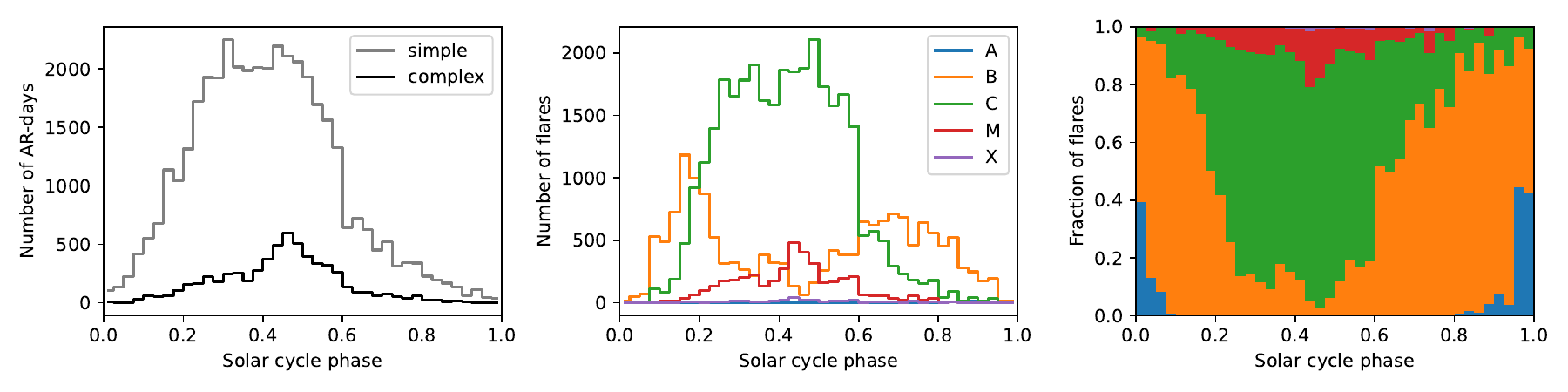}
\caption{Solar activity as a function of normalized solar-cycle phase. \textbf{Left:} Number of active-region days for magnetically simple and complex regions. \textbf{Center:} GOES flare counts by class. \textbf{Right:} Fractional GOES-class composition in each phase bin, normalized within each bin.}
\label{fig:phase}
\end{figure*}

\subsection{Cycle variations}
The GOES data are plotted as a function of solar-cycle phase, with the 11-year cycle \citep{Hathaway2010} normalized to span the range from 0 to 1. Each cycle's start and end times were defined by the dates given by the Royal Observatory of Belgium\footnote{\url{https://www.sidc.be/SILSO/cyclesminmax}}.

The first panel of Fig.~\ref{fig:phase} shows the frequency of simple and complex ARs over the course of the cycle. A clear asymmetry is seen in the occurrence of simple spots earlier on in the cycle, while more complex spots become more frequent towards the peak and later stages of the cycle. This is believed to be a result of so-called active longitudes, which require existing spots to be intermixed with emerging flux to form more complex active regions \citep[e.g.][]{Pevtsov2025,Finley2025,Kontogiannis2025}. 

This tracks well with the second and third panels of Fig.~\ref{fig:phase} where stronger flares follow a similar trend, and are clearly tied to the occurrence of complex ARs. Additionally, a clear bias is seen in the appearance of A- and B-class flares, which occur more prominently during the solar minimum. This is an observational bias caused by the fact that the GOES X-ray background rises above the B-class threshold during the solar maximum, thus drowning out the weaker flares. 

\subsection{Flare productivity}
As the SRS data are published daily, the evolution of sunspot groups is tracked at this cadence. Since ARs form, evolve, and decay continuously while only half the solar disk is visible at any given time, it is impractical to follow individual regions across their full lifetime. For this reason, each AR entry is treated as an independent observation for that day, expressed in the unit of AR-days. Thus, a single physical region is counted once for every day it appears in the SRS, contributing on average 7.1 AR-days per NOAA region in our sample. 

\begin{figure*}
\centering
\includegraphics[width=\textwidth]{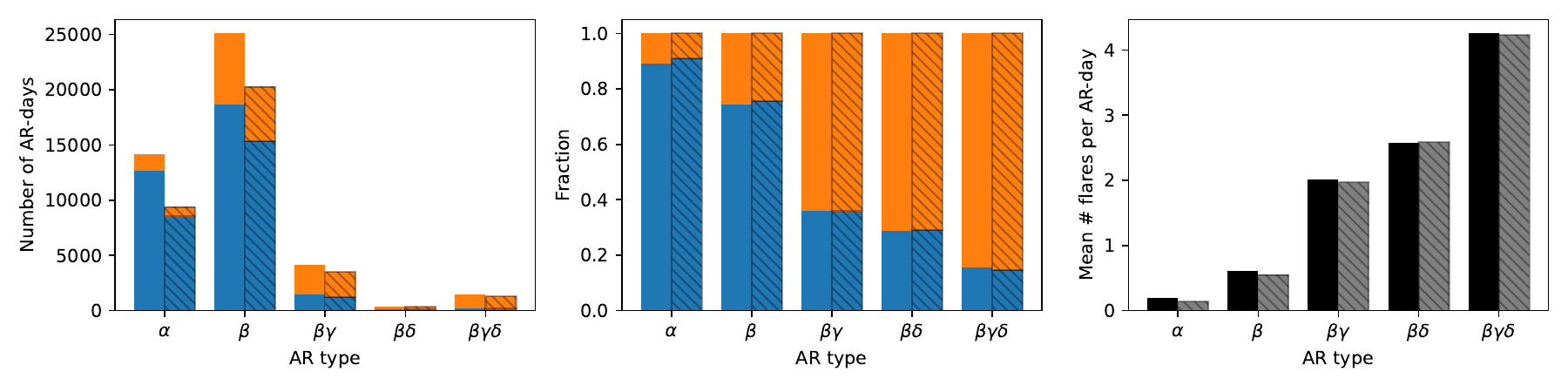}
\caption{Overview of flare productivity per AR type. Solid bars represent the full-disk sample, and the hatched bars correspond to the longitude-restricted sample. \textbf{Left:} Number of AR-days by Hale class, split into flaring (orange) and non-flaring (blue) populations. \textbf{Center:} Mean number of flares per AR-day. \textbf{Right:} Flaring fraction normalized within each Hale class. 
}
\label{fig:projection}
\end{figure*}

Although the SRS sunspot areas are already corrected for foreshortening, the Hale and Zurich/McIntosh classifications may still be harder to determine reliably near the limb due to reduced visibility of magnetic structure. To test the robustness of these classifications, a subsample was taken which was restricted to ARs within 60 degrees of the disk center, to avoid strongly projected ARs near the limb. In Fig.~\ref{fig:projection} the  number of ARs in each class are given for the whole sample (solid) and the subsample (dashed). In the first two panels, the ratio of flaring and non-flaring ARs is shown on a numerical and fractional scale, while the last panel shows the mean number of flares one can expect per AR-day. 

This figure demonstrates that projection effects do not introduce a significant bias to the spot classification, and thus do not argue against using the full dataset. It also shows that complex regions make up the bulk of the flare production, despite comprising only a small fraction of the total AR population. 

\begin{figure*}
\centering
\includegraphics[width=\textwidth]{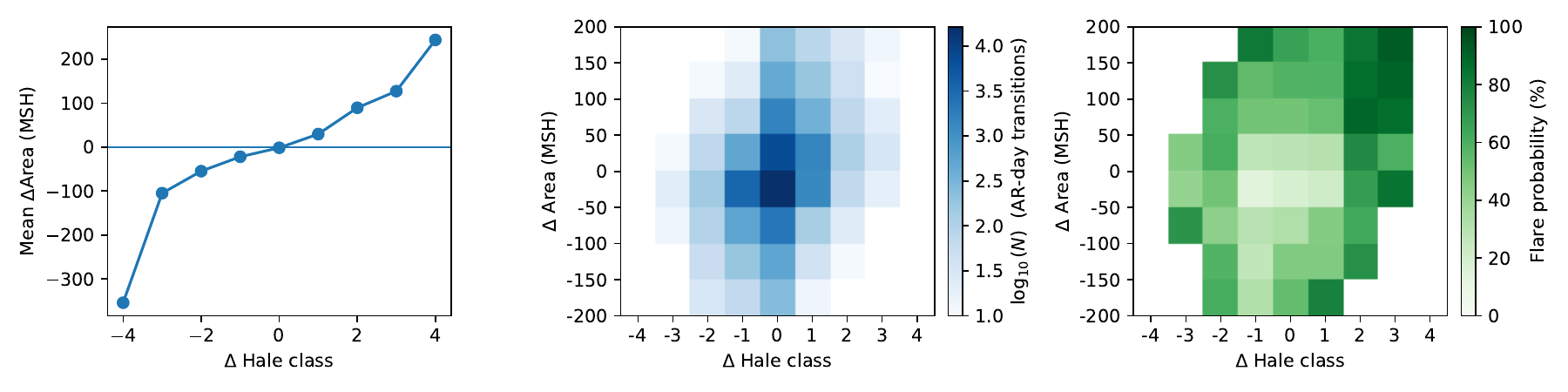}
\caption{Active region evolution and flare occurrence. \textbf{Left:} Mean change in sunspot area, \textbf{in bins of 50 MSH,} as a function of the change in Hale rank between consecutive days. \textbf{Center:} Number of AR-days in each bin (log scale). \textbf{Right:} Flare probability as a function of daily change in Hale rank and sunspot area. }
\label{fig:evolution}
\end{figure*}

In Fig.~\ref{fig:evolution}, flare productivity is mapped as a function of mean deprojected spot area, using 50 MSH bins, and changes in Hale class from one day to the next. Here, a change in Hale class is defined as the difference in complexity rank between consecutive days, such that a $\beta\gamma\delta$ region simplifying to an $\alpha$ region corresponds to a $\Delta\,\mathrm{Hale\,Class}$ of $-4$, and vice versa. The first panel shows an approximately linear relation between changes in Hale class and changes in area, with a steeper gradient at the extremes. The second panel shows the occurrence rate of each AR type, indicating that regions tend to maintain their Hale class but can change in area, and that large changes are rare over the course of a day. Whether this reflects a genuine physical tendency or an observational bias, whereby decaying regions receive less scrutiny from forecasters and are consequently updated less carefully in the SRS, remains unclear. However, these findings are in line with \citet{KILCIK2017}.

The final panel of Fig.~\ref{fig:evolution} shows the flare probability across this parameter space, defined as the fraction of AR-day-to-day transitions in each ($\Delta$ Hale rank, $\Delta$ Area) bin with at least one flare that day. Bins with n < 5 are masked to avoid unreliable small-number estimates. In general, flaring is most strongly associated with increasing Hale complexity, with a weaker tendency toward increasing area, although an additional peak appears for ARs that grow in area while decreasing in class, which contains comparatively more C-class flares and fewer M-class flares than regions growing in complexity. Regions that show little or no evolution are comparatively unlikely to flare. This is in line with the findings of \citet{Sammis2000}.

\begin{figure*}
\centering
\includegraphics[width=\textwidth]{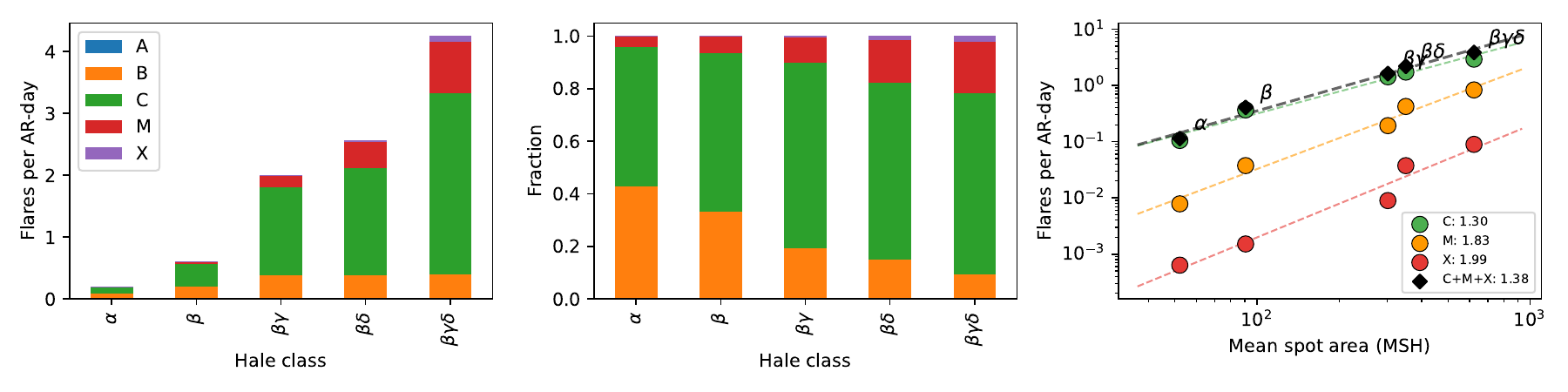}
\caption{Overview of flare productivity per flare type. \textbf{Left:} Flare productivity (flares per AR-day) by Hale class, stacked by GOES class. \textbf{Center:} Normalized GOES-class composition per Hale class. \textbf{Right:} Flare productivity as a function of mean sunspot area for C-, M-, and X-class flares, with power-law fits (dashed lines). The fitted exponents are given in the legend.}
\label{fig:productivityclass}
\end{figure*}

\begin{figure*}
\centering
\includegraphics[width=\textwidth]{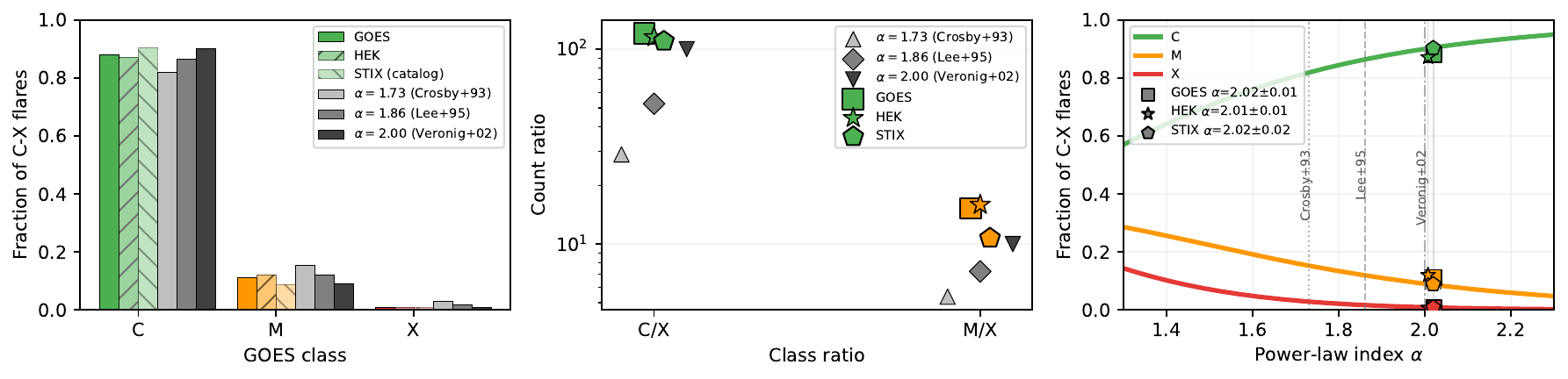}
\caption{Observed vs.\ predicted flare-class distributions. 
\textbf{Left:} Flare-class fractions normalised to the total C-X population for the GOES catalogue (solid bars), HEK (hatched with forward slashes), and STIX catalogue (hatched with backslashes), compared with predictions at three literature values of $\alpha$. X-class flares are restricted to X1--X9.9 to ensure equal sized bins. \textbf{Centre:} Observed C/X and M/X count ratios for each dataset compared with the same three literature predictions (grey symbols). \textbf{Right:} Predicted class fractions as continuous functions of $\alpha$ (solid curves), with each dataset plotted at its best-fit $\alpha$ obtained from a log-space minimisation.}
\label{fig:powerlaw}
\end{figure*}

Fig.~\ref{fig:productivityclass} shows the class-dependent flare productivity per AR type in units of flares per AR-day. It reproduces the same bias seen in Fig.~\ref{fig:phase}, with B-class flares becoming less frequent in more complex regions. Since this behavior is not seen for C-class flares and above, these higher classes allow a more direct comparison between flare productivity and the mean spot area of each AR type. 

A power-law relation between spot area and flare frequency was previously found by \citet{Eren2017} using the Zurich classification \citep{Waldmeier1955}, although no class subdivision was considered. In the final panel of Fig.~\ref{fig:productivityclass}, the same analysis is carried out for the full population as well as separately for C-, \mbox{M-,} and X-class flares. The resulting slopes become steeper with increasing flare class, indicating that more energetic flares are progressively more frequent in larger active regions. The strong resemblance between the C-class trend and that of the total flare population further shows that C-class flares dominate the overall distribution and, by extension, the fitted relations.

\subsection{Power-law estimation}\label{sec:powerlaw}

The solar flare frequency distribution as a function of the GOES peak flux has been a matter of discussion over the last few decades, as knowing it does not only help predict the next superflare \citep{Vasilyev2024}, but also helps us understand the size of the contribution to the energy budget of micro- and nanoflares towards the heating of the corona. The flare frequency is most typically expressed as a power-law based on the GOES peak flux, following the trend of
\begin{equation}
    \frac{dN}{dE} \propto F^{-\alpha},
    \label{eq:law}
\end{equation}
with $\alpha$ as the slope in peak flux, which serves as a proxy for total flare energy, and thus a proportionality constant between strong and weak flares. \citet{Hudson1991} showed that for $\alpha$ values of 2 or above, the energy budget of micro- and nanoflares would be sufficient to sustain coronal heating without requiring an additional mechanism. Naturally, the bulk of the literature swings around this value with $\alpha$ ranging between $1.70$ and $2.40$ as shown in Table~\ref{tab:alpha_lit}.

\begin{table}[h]
\caption{Literature values of the flare frequency power-law index $\alpha$.}
\label{tab:alpha_lit}
\centering
\footnotesize
\begin{tabular}{ll}
\toprule
Study & $\alpha$ \\
\midrule
\citet{Drake1971}      & $1.75 \pm 0.10$ \\
\citet{Crosby1993}     & $1.73 \pm 0.01$ \\
\citet{Lee1995}        & $1.86 \pm 0.10$ \\
\citet{Feldman1997}    & $1.88 \pm 0.21$ \\
\citet{Shimojo1999}    & $1.7 \pm 0.4$ \\
\citet{Veronig2002}    & $2.11 \pm 0.13$ \\
\citet{Veronig2002}$^{a}$ & $2.03 \pm 0.09$ \\
\citet{Yashiro2006}    & $2.11 \pm 0.03$ \\
\citet{Aschwanden2012} & $1.98 \pm 0.11$ \\
\citet{Ryan2016}$^{b}$ & $1.92$--$2.02$ \\
\citet{Hudson2024}     & $1.973 \pm 0.014$ \\
\citet{Biasiotti2025}$^{c}$ & $1.69$--$2.40$ \\
\bottomrule
\end{tabular}
\vspace{3pt}

\raggedright\footnotesize
$^{a}$background-subtracted. $^{b}$threshold-dependent. $^{c}$range across last four solar cycles.
\end{table}

In more recent years, the appropriateness of a simple power law has itself been questioned. \citet{Verbeeck2019} suggested that background-subtracted GOES peak flux data are better described by a lognormal distribution, while raw (uncorrected) data remain consistent with a power law. \citet{Hudson2024} instead suggested a tapered power law to account for lower occurrences of high-energy flares. \citet{Drake1971} and \citet{Biasiotti2025} noted that different energy bands seemingly have different slopes. In Fig.~\ref{fig:powerlaw}, this is explored by looking at the contributions of the C-, M-, and X-class flares, as both fractions and individual populations. To maintain equally sized energy bins, the X-class flares are limited to values below X10.

The first panel of Fig.~\ref{fig:powerlaw} shows the normalized fraction of the GOES and HEK data, as well as the newly released STIX flare catalogs, along with predicted values from the three aforementioned power-laws. The three flare catalogs are broadly consistent, with a slight skew towards weaker C-class flares from the STIX catalog. For the observational data, the C/X and M/X ratios are simply the count ratios between the respective classes. For a power law with index $\alpha$, the predicted ratios follow 
analytically from integrating Equation~\ref{eq:law} over the 
corresponding flux decades, yielding 
\begin{equation}
 \mathrm{C/X} = 
10^{2(\alpha-1)}  
\end{equation}
and
\begin{equation}
\mathrm{M/X} = 10^{(\alpha-1)}.   
\end{equation}
If the distribution is a single power law, the $\alpha$ implied by both ratios should agree. For the GOES and HEK datasets however, the M/X ratio implies a steeper slope ($\alpha = 2.18 \pm 0.03$ and $2.20 \pm 0.02$, respectively) than the C/X ratio ($\alpha = 2.04 \pm 0.01$ and $2.03 \pm 0.01$), yielding discrepancies of $\Delta\alpha = -0.14 \pm 0.03$ and $-0.17 \pm 0.02$, significant at the $4.5\sigma$ and $8\sigma$ level respectively. This indicates that the distribution steepens towards higher energies, with fewer X-class flares than a single power law extrapolated from the C- and M-class population would predict. The STIX data show no significant discrepancy ($\Delta\alpha = -0.01 \pm 0.05$), possibly reflecting the smaller sample size, sensitivity issues, or calibration bias.

One single power law index can also be inferred with the same 
class fractions by minimizing the predicted and observed fractions, as shown in the third panel of Fig.~\ref{fig:powerlaw}. This plot shows the power-law-predicted fraction of C-, M-, and X-class flares (solid curves) as a continuous function of $\alpha$, with each dataset plotted at its own best-fit value. The curves are asymmetric, with the C-class fraction staying much more constant than the stronger flares. This means the rarest, most energetic class is by far the most sensitive diagnostic of $\alpha$, but also the one most susceptible to small-number statistics, a property directly relevant to the C/X-versus-M/X discrepancy discussed above, since a deficit in X-class flares is precisely what would produce a steeper M/X-implied slope without much affecting the C/X-implied one.

However, given that the vast majority of all flares are of the C-class, a logarithmic fit is performed to prevent the C-class fraction from dominating the minimization. Uncertainties on the best-fit $\alpha$ 
are obtained by bootstrap resampling where $10\,000$ Poisson realizations of the observed counts are drawn, and $\alpha$ is recomputed for each realization. The standard deviation of the resulting distribution is taken as the $1\sigma$ uncertainty.

Using this joint log-space fraction fit, rather than the individual C/X or M/X ratios discussed above, all three datasets yield consistent power-law indices: $\alpha = 2.020 \pm 0.012$ (GOES), $2.007 \pm 0.008$ (HEK), and $2.019 \pm 0.018$ (STIX). These values cluster tightly around $\alpha = 2.0$, in agreement with \citet{Aschwanden2012} and \citet{Hudson2024}, and right at the critical threshold identified by \citet{Hudson1991} for nanoflare-driven coronal heating.

The fact that we obtain a steeper M/X-implied slope than a C/X-implied slope is consistent with a steepening of the flare frequency distribution towards higher energies (Fig.~\ref{fig:powerlaw}), in agreement with recent literature suggesting a tapered power law for the strongest flares \citep[e.g.,][]{Hudson2024}. This is perhaps best illustrated by extrapolating our fitted power-law index to the frequency of a Carrington-type superflare, estimated to have a strength of about X100 \citep{Carrington1859,Hudson2025}. Using the steeper, M/X-implied index ($\alpha \approx 2.18$), we obtain a recurrence time of about 26 years for such an event. This is clearly still an overestimate. 

We can make a similar cross-check using the flare productivity of the most magnetically complex ($\beta\gamma\delta$) active regions across the C, M, and X classes (2.94, 0.83, and 0.089 flares~AR-day$^{-1}$, respectively). Following \citet{Verbeeck2019} we fit these three points as a lognormal tail (Gaussian in $\log_{10}(F)$ with a peak flux $F_0 \approx 1.6\times10^{-7}$~W\,m$^{-2}$ and a width of 1.54~dex). Extrapolating this fit two further decades to the X10 and X100 ranges gives expected rates of $3.6\times10^{-3}$ and $5.6\times10^{-5}$ flares~AR-day$^{-1}$, respectively, translating to recurrence times of $5.6^{+1.9}_{-1.3}$~yr for an X10-class flare and $370^{+292}_{-159}$~yr for an X100-class event (median and 1$\sigma$ range from 10\,000 Poisson bootstrap resamples of the underlying flare counts). The X10 estimate is consistent with the six X10+ flares actually observed in our 30-year dataset, and the X100 estimate is compatible with the tapered-distribution estimates of \citet{Hudson2025} ($\sim$500~yr) and \citet{Elvidge2018} ($\sim$150~yr).

\section{Conclusions}

In this work, the flare productivity of solar active regions across solar cycles 23--25 was analyzed, using GOES event reports matched with Solar Region Summaries. The distribution of flare energies was also studied in combination with the 50-year HEK flare catalog and the recently released STIX flare list from Solar Orbiter. The main results are as follows:

\begin{enumerate}
    \item Complex active regions ($\beta\gamma$, $\beta\delta$, $\beta\gamma\delta$) produce the bulk of M- and X-class flares despite comprising only $\sim$13\% of all AR-days (see table~\ref{tab:hale_stats} and Fig.~\ref{fig:phase}). The flaring fraction and flare rate per AR-day increase monotonically with magnetic complexity (See Fig.~\ref{fig:projection}), and these trends are shown to not be affected by projection effects by comparing the full-disk sample to a subsample restricted to longitudes within 60$^\circ$ of disk center.
    \item Flare productivity is most strongly associated with increasing Hale complexity, with a secondary dependence on sunspot area (See Fig.~\ref{fig:evolution}). A power-law relation between mean spot area and flare rate steepens with GOES class, from $\beta = 1.30$ for C-class to $\beta = 1.99$ for X-class flares, indicating that the most energetic flares are disproportionately concentrated in the largest active regions (See Fig.~\ref{fig:productivityclass}).
    \item Three flare catalogs were used (GOES, HEK, 
    and STIX), with the last one being considered independent. All three yield a consistent power law index of $\alpha \approx 2.02 \pm 0.01$ for the peak flux frequency distribution, obtained via a log-space fit of the C-, M-, and X-class fractions. This value sits right at the critical threshold identified by \citet{Hudson1991} for nanoflare-driven coronal heating.
    However, the distribution is not well described by a single power law across all three classes. If fitted separately, a power law index of $\alpha\sim$2.2 is found for the M/X population, while a less steep value of $\alpha\sim$2.0 is found for the C/X ratio. This steepening at higher energies 
    is consistent with the tapered power-law behavior reported 
    by \citet{Hudson2024} and the cycle-dependent slope variations 
    found by \citet{Biasiotti2025}, and points to a decrease in frequency of super flares and potentially an upper limit for flare strength on the Sun.
    \item Extrapolating the observed steepening of flare productivity for the most magnetically complex ($\beta\gamma\delta$) active regions two further decades beyond the X-class threshold predicts recurrence times of $5.6^{+1.9}_{-1.3}$~yr for an X10-class flare and $370^{+292}_{-159}$~yr for an Carrington-like X100-class event. The X10 estimate is consistent with the six X10+ flares actually observed in our 30-year dataset, and the X100 estimate is considerably more compatible with independent tapered-distribution estimates \citep{Hudson2025, Elvidge2018} than a naive single power-law extrapolation, which instead predicts such an event once every $\sim$26~years.
\end{enumerate}

Together, these results provide a comprehensive statistical baseline for solar flare productivity as a function of active region properties, and highlight that while a power-law index of $\alpha \approx 2$ provides a useful global description, the underlying distribution exhibits structure that a single power law cannot fully capture.

\begin{acknowledgements}
AP was supported by grant PI~2102/1-1 from the Deutsche Forschungsgemeinschaft (DFG). I thank Alexander Warmuth for proofreading the manuscript, Henrik Eklund and Adam Finley for their discussion on this topic, and the anonymous referee for their feedback during the peer review process. DeepL Write was used for assistance with grammar and language polishing.
\end{acknowledgements}

\bibliographystyle{aa}
\bibliography{ref}

\end{document}